\documentclass[11pt,fleqn,a4paper]{article}
\usepackage{amsfonts,amssymb,amsmath}
\title{Hierarchy of Multipolar Terms for
Non Relativistic Moving Atoms in QED}
\author{\small }
\date{ \small {\bf Abdelbasset Felhi}\\
D\'{e}partement de Math\'{e}matiques, Facult\'{e} des Sciences de
Bizerte, Jarzouna-Bizerte, 7021 Tunisie,\\
 e-mail adress: Abdelbasset.Felhi@fsb.rnu.tn\\
 {\bf Jean-Claude Guillot}\\
CNRS-UMR 7539, D\'{e}partement de Math\'{e}matiques, Institut
Galil\'{e}e, Universit\'{e} Paris 13, F-93430 Villetaneuse and
CNRS-UMR 7641, Centre de Math\'{e}matiques Appliqu\'{e}es, Ecole
Polytechnique, F-91128 Palaiseau Cedex,\\ e-mail adress:
guillot@cmapx.polytechnique.fr\\ {\bf Jacques Robert} \\ CNRS-UMR
7538, Laboratoire de Physique des Lasers, Institut Galil\'{e}e,
Universit\'{e} Paris 13, F-93430 Villetaneuse,\\ e-mail adress:
robert@galilee.univ-paris13.fr}
\def\v{\varepsilon}
\def\pn{\par\noindent}

\newcommand{\er}{{\Bbb R}}

\begin{document}
\maketitle
\date
{\bf preprint} {QED 10/11/2003}\\
\begin{abstract}
In this article an hierarchy of terms in the expansion of the
multipolar Hamiltonian for non relativistic moving atoms in QED is
considered. The particular case of neutral composite systems of
2,3 and 4 particles is considered. The proposed hierarchy is based
on a scaling analysis of the multipolar Hamiltonian together with
a multipolar expansion of the coupling terms. We give explicit
results up to the fourth order in the fine structure constant.
\end{abstract}





\section{Introduction}

The theory of multipolar Hamiltonians for atoms in QED has been
mainly developped through the Power-Zienau-Wooley (PZW)
transformation for the minimal coupling Hamiltonian. This
formulation is well suited for the description of a confined
system of charges and is particularly relevant for a neutral
system where the vector potential disappears. For moving atoms
there have been few proposals of such Hamiltonians [H, LBBL,GR]
owing to the fact that, up to the first order in the fine
structure $\alpha,$ symetries of the $p\cdot A$ and $d\cdot E$
schemes are the same. Higher order terms has been discussed mostly
in the framework of two level systems [LBBL, W], and more recently
for real atoms [BBB]. The analysis of usually disregarded terms
[GR] appeals for a systematic procedure that can lead to an
unambiguous hierarchy of the terms in the multipolar expansions in
order to perform a relevant perturbation theory. In this note we
focus on QED of moving non relativistic light atoms, without
external fields. These atoms are composite systems of two, three
or four particles in order to make explicit the role of the
increasing complexity of the problem. The point is to examine how
the scaling technique performed from the beginning can be
associated with multipolar expansions in order to obtain a
straightforward hierarchy of terms in the Hamiltonian. The second
section reviews the QED multipolar Hamiltonian and the relevant
physical quantities, the third section presents a sketch of the
scaling technique and proposes a way to determine which terms are
predominant in the Hamiltonian and the fourth section makes a
direct use of that choice within the multipolar expansion of the
interaction terms and gives also the corresponding expansion for
the minimal coupling Hamiltonian. Finally, in the fifth section,
the use of Jacobi coordinates is introduced and explicit
expansions are given.\\

{\bf Acknowledgement}
\\
We thank M.T. Jaekel for valuable suggestions that have lead us to
a more systematic description of the proposed expansions.

\section{The multipolar Hamiltonian}

We consider a neutral system of $N$ charges $(e_{a})_{1\leq a\leq
N},$ with masses $(m_{a})_{1\leq a\leq N},$ spins $(S_{a})_{1\leq
a\leq N}$ and Land\'{e} coefficient $(g_{a})_{1\leq a\leq N}$. The
PZW Hamiltonian for QED
in Coulomb gauge is given by (see [GR]):%

\begin{align}
H_{mp}  &  =\sum_{a=1}^{N}\frac{1}{2m_{a}}\left(  p_{a}+\int\Theta
_{a}(r)\wedge B(r)\,d^{3}r\right)  ^{2}+\sum_{a<b}\frac{e_{a}e_{b}}%
{4\pi\varepsilon_{0}|r_{a}-r_{b}|}\nonumber\\ &
-\sum_{a=1}^{N}M_{a}.B(r_{a})+\frac{1}{2}\int d^{3}r~(\frac{(\Pi
(r)+P(r))^{2}}{\varepsilon_{0}}+\frac{B(r)^{2}}{\mu_{0}}), \label{1}%
\end{align}
with
\begin{equation}
p_{a}=-i\hbar\nabla_{r_{a}},\quad
M_{a}=g_{a}\frac{e_{a}}{2m_{a}}S_{a},\quad
c^{2}\varepsilon_{0}\mu_{0}=1\label{2}%
\end{equation}
where
\begin{align}
B(r)  &  =i\sum_{\lambda=1,2}\int d^{3}k\frac{1}{c}\left(
\frac{\hbar
\,\omega(k)}{2\varepsilon_{0}(2\pi)^{3}}\right)  ^{1/2}(a_{\lambda}%
(k)\,(\frac{k}{|k|}\wedge\varepsilon_{\lambda}(k))\,e^{ik\cdot
r}\nonumber\\ &
-a_{\lambda}^{\ast}(k)\,(\frac{k}{|k|}\wedge\varepsilon_{\lambda
}(k))\,e^{-ik\cdot r}) \label{3}%
\end{align}%
\begin{equation}
\Pi(r)=-i\sum_{\lambda=1,2}\int\left( \frac{\hbar\,\varepsilon
_{0}\,\omega(k)}{2(2\pi)^{3}}\right)
^{1/2}(a_{\lambda}(k)\,\varepsilon _{\lambda}(k)\,e^{ik\cdot
r}-a_{\lambda}^{\ast}(k)\,\varepsilon_{\lambda
}(k)\,e^{-ik\cdot r}) \label{4}%
\end{equation}%
\begin{equation}
\text{and\qquad\ }E(r)=-\frac{1}{\varepsilon_{0}}\Pi(r). \label{5}%
\end{equation}
Here $\omega(k)=c|k|$ and $\varepsilon_{\lambda}(k)$,
$\lambda=1,2$, are real polarization vectors satisfying
\begin{equation}
\varepsilon_{\lambda}(k)\cdot\varepsilon_{\mu}(k)=\delta_{\lambda\mu},\quad
k\cdot\varepsilon_{\lambda}(k)=0,\quad\lambda,\mu=1,2 \label{6}%
\end{equation}
$a_{\lambda}(k)$, $a_{\lambda}^{\ast}(k)$ are the usual
annihilation and creation operators obeying the canonical
commutation relations
\begin{equation}
\lbrack
a_{\lambda}^{\sharp}(k),a_{\mu}^{\sharp}(k)]=0,\quad\lbrack
a_{\lambda}(k),a_{\mu}^{\ast}(k^{\prime})]=\delta_{\lambda\mu}\delta
(k-k^{\prime}) \label{7}%
\end{equation}
where $a^{\sharp}=a$ or $a^{\ast}$.

The polarization field relative to the center-of-mass coordinate
$R$ is:
\begin{equation}
P(r)=\sum_{a=1}^{N}e_{a}(r_{a}-R)\int_{0}^{1}d\lambda\,\delta(r-R_1-\lambda
(r_{a}-R)) \label{8}%
\end{equation}

The magnetization fields relative to the center-of-mass coordinate
$R,$ for
$1\leq$a$\leq N$ are%

\begin{equation}
\Theta_{a}(r)=\sum_{b=1}^{N}e_{b}(r_{b}-R)\int_{0}^{1}d\lambda(\lambda
\delta_{ab}-\frac{m_{a}}{M}(\lambda-1))\,\delta(r-R_1-\lambda(r_{b}-R))
\label{9}%
\end{equation}
where $R=\frac{1}{M}\,\sum_{a=1}^{N}m_{a}r_{a}$ with
$M=\sum_{a=1}^{N}m_{a}$.

$\ $

\section{Scaling}

\subsection{General Remarks}

The scaling is a change of units which exhibits the pertubative
character of the different terms in the Hamiltonian that we are
concerned with. Here we use dilatation of particles coordinates
and photon momenta independently. Then we have to establish which
terms are to be predominant in the Hamiltonian of this composite
(particles and photons) system. These terms will form the zero
order Hamiltonian for the perturbative expansion. In a first step,
we have to determine a relevant truncated zero order Hamiltonian
extending to moving atoms the properties known for non moving
atoms.

In the hydrogenoid model of atoms with independent electrons,
typical binding energy for each electron is of the order of
$\frac{1}{2}mc^{2}(Z\alpha)^{2}$ [M] where
$\alpha=\frac{e^{2}}{4\pi\varepsilon_{0}\hbar c}$ is the fine
structure constant, $Z$ the nucleus charge and $m$ the electron
mass. This quantity, which neglects all the further corrections
from inter electron repulsion sceening effects and exchange
process, is an upper value for the total ionisation of an atom.
Indeed the maximum atomic value for the first ionisation is that
of the helium atom: 24.58 eV, of the order of
$(\frac{1}%
{2}mc^{2}\alpha^{2}Z^{2})/2$. As the multipolar formalism fits at
best with confined system of charges we will disregard energies
exceeding the first ionisation limit. Thus it is consistent to
built a model where the Coulombic part, kinetic energy of the
electrons and the free field energy have to be retained within the
leading terms [FFG]. For the nuclear motion the question of the
scaling factor arises: should we choose the same for nuclei and
electrons? One point of view is to respect the consistency between
different choices of coordinate systems for N particles systems:
the natural one $\left\{  r_{a}\right\}_{a=[1,N]}$ and the Jacobi
ones $\left\{R_{a}\right\}_{a=[1,N]}$ which are generated by a
definite building up of the center of mass and relative vectors
and associated coordinates. As the constitutive set of equations
corresponding to the transformation from one system to the other
is a linear one, we have chosen to adopt a same scaling constant
for nuclei and electrons in order to fit with this property.

If the nucleus is the particle $a=1$, the unperturbed zero order
Hamiltonian is the following one:%

\begin{equation}
H_{0}=\sum_{a=1}^{N}\frac{p_a^2}{2m_{a}}+\sum
_{a=2}^{N}\frac{Ze^{2}}{4\pi\varepsilon_{0}|r_{a}-r_{1}|}+\frac{1}{2}\int
d^{3}r~(\frac{\Pi(r)^{2}}{\varepsilon_{0}}+\frac{B(r)^{2}}{\mu_{0}}),
\label{10}%
\end{equation}

where $N=Z+1$.

\subsection{Dilatation}

We dilate the particles coordinates and photon momenta
independently, \textit{i.e.,}
$(\{r_{a}\}_{a=[1,N]},k)\rightarrow\left( \{\frac{r_{a}}{\mu
}\}_{a=[1,N]},\eta k\right)  $, $\mu>0$, $\eta>0$, where $\mu$ and
$\eta$ are dimensionless constants. This scaling induces a unitary
transformation in the space of physical states that we now
compute.

The dilation $k\rightarrow\eta k$ induces an unitary operator in
$L^{2}(I\!\!R^{3})$ denoted by $\Gamma_{\eta}$ :
\begin{equation}
(\Gamma_{\eta}f)\,(k)=\eta^{3/2}f(\eta k),\qquad\eta>0,\quad k\in
I\!\!R^{3}
\label{11}%
\end{equation}
where $f\in L^{2}(I\!\!R^{3})$.

We have for the inverse transformation
\begin{equation}
(\Gamma_{\eta}^{-1}g)\,(k)=\eta^{-3/2}\,g\left(
\frac{k}{\eta}\right)
\label{12}%
\end{equation}

Let $\mathcal{F}_{ph}$ be the Fock space for transversal photons.
\\
We have $\displaystyle\mathcal{F}_{ph}=\bigoplus_{n=0}^{\infty}
L^{2}(I\!\!R^{3},I\!\!\!\!C^{2}%
)^{\otimes_{s}^{n}},$ where
$L^{2}(I\!\!R^{3},I\!\!\!\!C^{2})^{\otimes_{s}^{0}%
}\equiv I\!\!\!\!C.$ Here
$L^{2}(I\!\!R^{3}%
,I\!\!\!\!C^{2})^{\otimes_{s}^{n}}$ is the symmetrized $n$-fold
tensor product of $L^{2}(I\!\!R^{3},I\!\!\!\!C^{2})$. The dilation
$k\rightarrow\eta k$, $\eta>0$, induces an unitary operator on
$\mathcal{F}_{ph}$ that we still denote by $\Gamma_{\eta}$. We
have for $\psi=(\psi^{(n)})_{n\geq0}\in\mathcal{F}$:
\begin{align}
(\Gamma_{\eta}\psi)^{n}(k_{1},\lambda_{1};\cdots;k_{n},\lambda_{n})
& =\eta^{\frac{3}{2}n}\psi^{(n)}(\eta
k_{1},\lambda_{1};\cdots;\eta k_{n},\lambda_{n}),\label{13}\\
k_{j}  &  \in I\!\!R^{3},\ \lambda_{j}=1,2,\ j=1,\cdots
n.\nonumber
\end{align}
Thus we have
\begin{align}
\Gamma_{\eta}^{-1}a_{\lambda}(k)\Gamma_{\eta}  &
=\eta^{3/2}a_{\lambda}(\eta k)\label{14}\\
\Gamma_{\eta}^{-1}a_{\lambda}^{\ast}(k)\Gamma_{\eta}  &  =\eta^{3/2}%
a_{\lambda}^{\ast}(\eta k) \label{15}%
\end{align}
because $a_{\lambda}^{\ast}(k)$ is the adjoint of
$a_{\lambda}(k)$.

The right mathematical definition of the energy operator for the
photons is given by
\begin{equation}
\frac{1}{2}\int
d^{3}r:\frac{\Pi(r)^{2}}{\varepsilon_{0}}+\frac{B(r)^{2}}%
{\mu_{0}}: \label{16}%
\end{equation}
where :\ : is the Wick normal order (see [S]).

Let $H_{f}$ denote the energy operator for the photons. We then
have :
\begin{equation}
H_{f}=\frac{1}{2}\int
d^{3}r:\frac{\Pi(r)^{2}}{\varepsilon_{0}}+\frac
{B(r)^{2}}{\mu_{0}}:=\sum_{\lambda=1,2}c\hbar\int
d^{3}k\,|k|a_{\lambda}%
^{\ast}(k)a_{\lambda}(k). \label{17}%
\end{equation}
Thus using (3), (4), (5), (19) and (14), (15) one gets:%

\begin{align}
\Gamma_{\eta}B(r)\Gamma_{\eta}^{-1}  &  =\eta^{2}B(\eta
r),\label{18}\\ \Gamma_{\eta}\Pi(r)\Gamma_{\eta}^{-1}  &
=\eta^{2}\Pi(\eta r),\label{19}\\
\Gamma_{\eta}E(r)\Gamma_{\eta}^{-1}  &  =\eta^{2}E(\eta
r),\label{20}\\
\Gamma_{\eta}H_{f}\Gamma_{\eta}^{-1}  &  =\eta H_{f}. \label{21}%
\end{align}

Moreover
\begin{equation}
\Gamma_{\mu}^{-1}\nabla_{r_{a}}\,\Gamma_{\mu}=\mu\,\nabla_{r_{a}}.
\label{22}%
\end{equation}
Furthermore, for any function $V(r_{a})$, we have
\begin{equation}
\Gamma_{\mu}^{-1}V(r_{a})\,\Gamma_{\mu}=V\left(
\frac{r_{a}}{\mu}\right).
\label{23}%
\end{equation}

We have to transform the multipolar Hamiltonian by the total
dilatation $\Gamma$ acting both on particles coordinates and
photon momenta:
$\Gamma=(\Gamma_{\frac{1}{\mu}})^{N}\Gamma_{\eta}$. Thus we now get:%

\begin{subequations}
\begin{align}
\Gamma B(r)\Gamma^{-1}&=\eta^{2}B(\frac{\eta r}{\mu}),\label{24}\\
\Gamma\Pi(r)\Gamma^{-1}&=\eta^{2}\Pi(\frac{\eta
r}{\mu}),\label{25}\\ \Gamma E(r)\Gamma^{-1}&=\eta^{2}E(\frac{\eta
r}{ \mu}),\label{26}\\ \Gamma H_{f}\Gamma^{-1}&=\eta
H_{f},\label{27}\\ \Gamma
p_{a}\,\Gamma^{-1}&=\mu\,p_{a},\label{28}\\ \Gamma
V(r_{a})\,\Gamma^{-1}&=V\left( \frac{r_{a}}{\mu}\right).
\label{29}%
\end{align}

\subsection{Determination of the common scaling constant}

We first examine the action of $\Gamma$ on the truncated Hamiltonian:%

\end{subequations}
\begin{align}
&\Gamma H_{0}\Gamma^{-1}=\nonumber\\ &\Gamma\left[
\sum_{a=1}^{N}\frac{1}{2m_{a}%
}\left(  p_{a}\right)
^{2}+\sum_{a=2}^{N}\frac{Ze^{2}}{4\pi\varepsilon
_{0}|r_{a}-r_{1}|}+\frac{1}{2}\int
d^{3}r~(\frac{\Pi(r)^{2}}{\varepsilon_{0}%
}+\frac{B(r)^{2}}{\mu_{0}})\right]\Gamma^{-1}\label{30}\\
&=\mu^{2}\sum_{a=1}^{N}\frac{1}{2m_{a}}\left(  p_{a}\right)
^{2}+Z\alpha \mu\sum_{a=2}^{N}\frac{\hbar
c}{|r_{a}-r_{1}|}+\eta\frac{1}{2}\int
d^{3}r~(\frac{\Pi(r)^{2}}{\varepsilon_{0}}+\frac{B(r)^{2}}{\mu_{0}})\nonumber
\end{align}

Choosing the kinetic energy of the electrons, the free energy of
the photons and the Coulombic part as the leading terms of the
same order we must
have:%

\begin{equation}
\mu^{2}=Z\alpha\mu=\eta\label{31}%
\end{equation}

We then get:%

\begin{equation}
\eta=(Z\alpha)^{2},\,\text{\quad}\mu=Z\alpha\text{\quad
and\quad}\frac{\eta}%
{\mu}=Z\alpha\label{32}%
\end{equation}

and we have:%

\begin{equation}
\begin{split}
&\Gamma H_{0}\Gamma^{-1}=\\
&\mu^{2}\left[\sum_{a=1}^{N}\frac{1}{2m_{a}}\left(
p_{a}\right)^{2}+\sum_{a=2}^{N}\frac{\hbar
c}{|r_{a}-r_{1}|}+\frac{1}{2}\int
d^{3}r~(\frac{\Pi(r)^{2}}{\varepsilon_{0}}+\frac{B(r)^{2}}{\mu_{0}})\right]\label{33}
\end{split}
\end{equation}
where the electronic ground state energy corresponding to the term
within brackets is $\frac{1}{2}mc^{2}Z$ as $m_{a}=m$ for
a$=[2,N]$.

\section{Scaled Multipolar expansion}

\subsection{Scaled multipolar Hamiltonian}

The scaling conditions $(\ref{24}-\ref{29},\,\ref{32})$ and the
equations $(\ref{8}-\ref{9})$ will now allow us to dilate all the
terms of the multipolar Hamiltonian given by equation $(\ref{1}),$
once the expansion of squared bracket terms is performed.

%

\begin{align}
H_{mp}&=\sum_{a=1}^{N}\frac{1}{2m_{a}}p_{a}^{2}-%
{\displaystyle\sum\limits_{a=2}^{N}}
\frac{Ze^{2}}{4\pi\varepsilon_{0}|r_{a}-r_{1}|}+\frac{1}{2}\int d^{3}%
r~(\frac{\Pi(r)^{2}}{\varepsilon_{0}}+\frac{B(r)^{2}}{\mu_{0}})\nonumber\\
&+%
{\displaystyle\sum\limits_{2\leq a<b}}
\frac{e^{2}}{4\pi\varepsilon_{0}|r_{a}-r_{b}|}+\int
d^{3}r~\frac{\Pi (r).P(r)}{\varepsilon_{0}}\nonumber\\
&+\sum_{a=1}^{N}\frac{1}{m_{a}}p_{a}.\int
d^{3}r~\Theta_{a}(r)\wedge
B(r)-\sum_{a=1}^{N}M_{a}.B(r_{a})\label{34}\\
&+\sum_{a=1}^{N}\frac{1}{2m_{a}}(\int d^{3}r~\Theta_{a}(r)\wedge
B(r)\,)^{2}+\frac{1}{2~\varepsilon_{0}}\int
d^{3}r~P(r)^{2}\nonumber
\end{align}

\bigskip
Constrains (\ref{31}) and (\ref{32}) fix the following behaviour
for the
involved operators:%

\begin{align}
\Gamma H_{f}\Gamma^{-1}&=\mu^{2}H_{f},\label{35}\\ \Gamma
B(r)\Gamma^{-1}  &  =\mu^{4}B(\mu r),\label{36}\\
\Gamma\Pi(r)\Gamma^{-1}  &  =\mu^{4}\Pi(\mu r),\label{37}\\ \Gamma
E(r)\Gamma^{-1}  &  =\mu^{4}E(\mu r)\label{38}.
\end{align}

Now it remains to compute the effect of the scaling transformation
on terms including $P(r)$ and $\Theta_{a}(r):$

\begin{center}%
\begin{align}
&\Gamma\int d^{3}r~\frac{\Pi(r).P(r)}{\varepsilon_{0}}\,\Gamma
^{-1}\nonumber\\
&=\mu^{3}\,\sum_{a=1}^{N}e_{a}(r_{a}-R).\int_{0}^{1}d\lambda\int
d^{3}r~\frac{1}{\varepsilon_{0}}\Pi(r)~\delta(r-\mu\lbrack
R-\lambda (r_{a}-R)])\nonumber\\ &
=\mu^{3}\int_{0}^{1}d\lambda~\sum_{a=1}^{N}e_{a}(r_{a}-R).\frac
{1}{\varepsilon_{0}}\Pi(\mu\lbrack
R-\lambda(r_{a}-R)]),\label{41}\\ &\Gamma\int
d^{3}r~\Theta_{a}(r)\wedge B(r)\Gamma^{-1}=
\mu^{3}\sum_{b=1}^{N}e_{b}(r_{b}-R)\wedge\int_0^1
(\lambda\delta_{ab}-\frac{m_{a}}{M}(\lambda-1))\nonumber\\
 &\int
d^{3}r~B(r)~\delta (r-R_1-\lambda(r_{b}-R))\nonumber\\
&=\mu^{3}\,\int_{0}^{1}d\lambda\sum_{b=1}^{N}(\lambda\delta_{ab}-\frac
{m_{a}}{M}(\lambda-1))e_{b}(r_{b}-R)\wedge\,~B(\mu\lbrack
R-\lambda
(r_{b}-R)]). \label{42}%
\end{align}

\end{center}

\bigskip

Finally one obtains for the multipolar Hamiltonian the
following scaling:%

\begin{align}
&\frac{1}{\mu^{2}}\Gamma H_{mp}\Gamma^{-1}=\sum_{a=1}^{N}\frac{1}{2m_{a}%
}p_{a}^{2}-%
{\displaystyle\sum\limits_{a=2}^{N}}
\frac{\hbar c}{|r_{a}-r_{1}|}+\frac{1}{2}\int d^{3}r~(\frac{\Pi(r)^{2}%
}{\varepsilon_{0}}+\frac{B(r)^{2}}{\mu_{0}})\nonumber\\
&+\frac{1}{Z}%
{\displaystyle\sum\limits_{2\leq a<b}}
\frac{\hbar c}{|r_{a}-r_{b}|}+\mu\int_{0}^{1}d\lambda~\sum_{a=1}^{N}%
e_{a}(r_{a}-R).\frac{1}{\varepsilon_{0}}\Pi(\mu\lbrack R-\lambda
(r_{a}-R)])\nonumber\\
&-\mu^{2}\sum_{a=1}^{N}M_{a}.B(\mu~r_{a})\label{43}\\
&+\mu^{2}\int_{0}^{1}d\lambda\sum_{a,b=1}^{N}(\lambda\delta_{ab}-\frac
{m_{a}}{M}(\lambda-1))\frac{e_{b}}{m_{a}}\nonumber\\
&\left(
p_{a}.\left((r_{b}-R)\wedge B(\mu\lbrack
R-\lambda(r_{b}-R)])\right)\right)\nonumber\\
&+\mu^{4}\sum_{a=1}^{N}\frac{1}{2m_{a}}(\int_{0}^{1}d\lambda\sum_{b=1}%
^{N}(\lambda\delta_{ab}-\frac{m_{a}}{M}(\lambda-1))e_{b}\nonumber\\
&(r_{b}-R)\wedge B(\mu\lbrack
R-\lambda(r_{b}-R)])\,)^{2}\nonumber\\
 &=\frac{1}{\mu^2}\Gamma H_0\Gamma^{-1}+\frac{1}{\mu^2}\Gamma
H_{mp}^I\Gamma^{-1}. \nonumber
\end{align}
As usual the vacuum polarization term, which scales as $\mu^{-2}$
has been forgetten in $(\ref{43}),$ and the last equality defines
$H_{mp}^I.$
\subsection{Multipolar expansions}

The multipolar expansion of the expression (\ref{43}) is
straightforward. It
relies on the usual Taylor series expansion for $F=B$ or $\Pi$:%
\begin{equation}
F(\mu\left[  R-X\right]
)=\sum_{n=0}^{\infty}\mu^{n}\,\frac{\left(
X.\nabla_{r}\right)  ^{n}}{n!}F(r)_{\mid r=\mu R} \label{44}%
\end{equation}
where $X$ stands for $(r-R_{a})$ or $\lambda(r_{a}-R)~$.

Note that $\left\vert k\right\vert \left\vert r_{\alpha
}-R\right\vert $ is of the order of $|k|a_{1}(Z)$ where
$a_{1}(Z)=mcZ\alpha /\hbar$. As the photons energy has to be lower
to the value $mc^2$ which is used for the ultraviolet cutoff
[BFS], it then results that $|k|<mc/\hbar$. Therefore the
$\mu\left\vert k.(r_{a}-R)\right\vert $ factor in the exponential
terms of the field expansions (\ref{3}) and (\ref{4}) can be
assumed to be small (mod $2\pi$).

\bigskip

One then gets the following explicit $\mu-$expansions for these
terms.

\subparagraph{\bigskip Electric multipole terms:}

\begin{equation}
\mu\int_{0}^{1}d\lambda~\sum_{\alpha=1}^{N}e_{\alpha}(r_{\alpha}-R).\frac
{1}{\varepsilon_{0}}\Pi(\mu\lbrack
R-\lambda(r_{\alpha}-R)])\nonumber
\end{equation}%
\begin{equation}
=\sum_{n=0}^{\infty}\mu^{n+1}\sum_{\alpha=1}^{N}e_{\alpha}(r_{\alpha}%
-R).\frac{\left(  (r_{\alpha}-R).\nabla_{r}\right)
^{n}}{(n+1)!}\frac
{\Pi(r)_{\mid r=\mu R}}{\varepsilon_{0}} \label{46}%
\end{equation}%
\[
=\sum_{n=0}^{\infty}\mu^{n+1}T_{E}^{n}(\mu)%
\]

\subparagraph{Spin mutipole terms}%

\[
\mu^{2}\sum_{\alpha=1}^{N}M_{\alpha}.B(\mu~r_{\alpha})
\]%
\begin{equation}
=\sum_{n=0}^{\infty}\mu^{n+2}(-1)^{n}\sum_{\alpha=1}^{N}M_{\alpha}%
.\frac{\left(  (r_{\alpha}-R).\nabla_{r}\right)
^{n}}{n!}B(r)_{\mid
r=\mu R} \label{47}%
\end{equation}%
\[
=\sum_{n=0}^{\infty}\mu^{n+2}T_{S}^{n}(\mu)%
\]

\subparagraph{Magnetic multipole terms}%
\begin{align}
&\mu^{2}\int_{0}^{1}d\lambda\sum_{\alpha,\beta=1}^{N}(\lambda\delta
_{\alpha\beta}-\frac{m_{\alpha}}{M}(\lambda-1))\frac{e_{\beta}}{m_{\alpha}}\nonumber\\
&\left( p_{\alpha}.\left((r_{\beta}-R)\wedge B(\mu\lbrack
R-\lambda (r_{\beta}-R)])\right)\right)\nonumber\\
&=\sum_{n=0}^{\infty}\mu^{n+2}\{\int_{0}^{1}d\lambda\sum_{\alpha,\beta
=1}^{N}\lambda^{n}(\lambda\delta_{\alpha\beta}-\frac{m_{\alpha}}{M}%
(\lambda-1))\nonumber\\
&  \frac{e_{\beta}}{m_{\alpha}}\left(  p_{\alpha}.\left(  (r_{\beta}%
-R)\wedge\frac{\left(  (r_{\alpha}-R).\nabla_{r}\right)  ^{n}}{n!}%
B(r)_{\mid r=\mu R}\right)  \right)  \} \label{48}%
\end{align}%
\[
=\sum_{n=0}^{\infty}\mu^{n+2}T_{M}^{n}(\mu)%
\]

\subparagraph{Diamagnetic multipole terms}%

\begin{align}
&\mu^{4}\sum_{\alpha=1}^{N}\frac{1}{2m\alpha}(\int_{0}^{1}d\lambda\sum
_{\beta=1}^{N}(\lambda\delta_{\alpha\beta}-\frac{m_{\alpha}}{M}(\lambda
-1))e_{\beta}\nonumber\\ &(r_{\beta}-R)\wedge B(\mu\lbrack
R-\lambda(r_{\beta}-R)])\,)^{2}\nonumber\\
&=\mu^{4}\sum_{\alpha=1}^{N}\frac{1}{2m\alpha}\{\int_{0}^{1}d\lambda
\sum_{\beta=1}^{N}\sum_{n=0}^\infty\nonumber\\
&\mu^{n}\lambda^{n}(\lambda\delta_{\alpha\beta}%
-\frac{m_{\alpha}}{M}(\lambda-1))e_{\beta}\left(
(r_{\beta}-R)\wedge \frac{\left(  (r_{\alpha}-R).\nabla_{r}\right)
^{n}}{n!}B(r)_{\mid r=\mu R}\right)\}^{2} \label{49}\\
&=\sum_{n=0}^{\infty}\mu^{n+4}T_{MM}^{n}(\mu)\nonumber
\end{align}

This allows to perform direct expansion up to the required order
in $\mu=Z\alpha$ of the multipolar Hamiltonian for an atom
considered as a confined neutral system of charges. Assuming that
$T_{E}^{n},T_{S}^{n},T_{M}^{n}%
,T_{MM}^{n}=0$ for $n<0$, one gets for the interaction Hamiltonian
$H_{mp}^{I}$ the following multipolar expansion:%
\begin{equation}
\frac{1}{\mu^2}\Gamma
H_{mp}^{I}\Gamma^{-1}=\sum_{n=0}^{\infty}\mu^{n}\left\{
T_{E}^{n-1}(\mu)+T_{S}^{n-2}(\mu)%
+T_{M}^{n-2}(\mu)+T_{MM}^{n-4}(\mu)\right\}  \label{50}%
\end{equation}
\subsection{Multipolar expansion of the minimal coupling Hamiltonian.}

For sake of completeness let us remind the expression of the
minimal coupling
Hamiltonian corresponding to the same system of charges as in (\ref{1}):%

\begin{align}
H_{mc}  &  =\sum_{\alpha=1}^{N}\frac{1}{2m_{\alpha}}\left(
p_{\alpha }-e_{\alpha}A(r)\right)
^{2}+\sum_{\alpha<\beta}\frac{e_{\alpha}e_{\beta}%
}{4\pi\varepsilon_{0}|r_{\alpha}-r_{\beta}|}\nonumber\\
&  -\sum_{\alpha=1}^{N}M_{\alpha}.B(r_{\alpha})+\frac{1}{2}\int d^{3}%
r~(\frac{\Pi(r)^{2}}{\varepsilon_{0}}+\frac{B(r)^{2}}{\mu_{0}}),
\label{51}%
\end{align}
where
\begin{equation}
A(r)=\sum_{\lambda=1,2}\int d^{3}k\left(
\frac{\hbar}{2\epsilon_{0}\left( 2\pi\right)
^{3}\omega(k)}\right)  ^{\frac{1}{2}}\varepsilon_{\lambda
}(k)\left\{
a_{\lambda}(k)e^{ik.r}+a_{\lambda}^{\ast}(k)e^{-ik.r}\right\}
\label{52}%
\end{equation}

One easily computes that $\Gamma A(r)\Gamma^{-1}=\mu^{2}A(\mu r)$.
For the other terms the preceding analysis remains inchanged. As a
result one gets a
scaled expression for the minimal coupling Hamiltonian:%

\begin{align}
\frac{1}{\mu^{2}}\Gamma H_{mc}\Gamma^{-1}  &
=\sum_{\alpha=1}^{N}\frac
{1}{2m_{\alpha}}p_{\alpha}^{2}-%
{\displaystyle\sum\limits_{\alpha=2}^{N}}
\frac{\hbar c}{|r_{\alpha}-r_{1}|}+\frac{1}{2}\int
d^{3}r~(\frac{\Pi(r)^{2}%
}{\varepsilon_{0}}+\frac{B(r)^{2}}{\mu_{0}})\nonumber\\
&  +\frac{1}{Z}%
{\displaystyle\sum\limits_{2\leq\alpha<\beta}}
\frac{\hbar
c}{|r_{\alpha}-r_{\beta}|}-\mu\sum_{\alpha=1}^{N}\frac{e_{\alpha}%
}{m_{\alpha}}p_{\alpha}.A(\mu r_{\alpha})\nonumber\\ &
-\mu^{2}\sum_{\alpha=1}^{N}M_{\alpha}.B(\mu~r_{\alpha})+\mu^{2}\sum
_{\alpha=1}^{N}\frac{e_{\alpha}^{2}}{2m_{\alpha}}A(\mu
r_{\alpha})^{2}\nonumber\\ &=\frac{1}{\mu^2}\Gamma
H_0\Gamma^{-1}+\frac{1}{\mu^2}H_{mc}^{I}\Gamma^{-1},
\label{53}%
\end{align}
where the right hand side defines $H_{mc}^{I}.$

The multipolar expansion is the consequence of (\ref{44}) for
$F=A$ and
$X=(r-R_{a})$:%

\begin{equation}
\begin{split}
&\mu\sum_{\alpha=1}^{N}\frac{e_{\alpha}}{m_{\alpha}}p_{\alpha}.A(\mu
r_{\alpha })\\
&=\sum_{n=0}^{\infty}\mu^{n+1}(-1)^{n}\sum_{\alpha=1}^{N}\frac{e_{\alpha}%
}{m_{\alpha}}p_{\alpha}.\frac{\left(
(r_{\alpha}-R).\nabla_{r}\right)^{n}}{n!}A(r)_{\mid r=\mu R}\\ &
=\sum_{n=0}^{\infty}\mu^{n+1}T_{A}^{n}(\mu)\label{54}.
\end{split}
\end{equation}

and%

\begin{equation}
\begin{split}
&\mu^{2}\sum_{\alpha=1}^{N}\frac{e_{\alpha}^{2}}{2m_{\alpha}}A(\mu
r_{\alpha})^{2}=\sum_{n=0}^{\infty}\mu^{n+2}\times\\
&\sum_{l=0}^{n}\sum_{\alpha=1}^{N}\frac{e_{\alpha
}^{2}}{2m_{\alpha}}\frac{((r_{\alpha}-R).\nabla_{r})^{n-l}}{(n-l)!}A(r)_{\mid
r=\mu R}.\frac{( (r_{\alpha}-R).\nabla_{r})^{l}}{l!}A(r)_{\mid
r=\mu R}\\
&=\sum_{n=0}^{\infty}\mu^{n+2}T_{AA}^{n}(\mu).\label{55}
\end{split}
\end{equation}

Assuming that $T_{A}^{n},\,T_{AA}^{n}=0$ for $n<0$, one gets for
the minimal
coupling interaction Hamiltonian the following expansion:%
\begin{equation}
\frac{1}{\mu^2}\Gamma
H_{mc}^{I}\Gamma^{-1}=\sum_{n=0}^{\infty}\mu^{n}\left\{
T_{A}^{n-1}(\mu)+T_{S}^{n-2}(\mu) +T_{AA}^{n-2}(\mu)\right\}
\label{56}
\end{equation}

Thus the correspondence between terms of the same order in $\mu$
for $H_{mp}$ and $H_{mc}$ can be established by direct comparison
of the expressions (\ref{43}) and (\ref{51}) or (\ref{50}) and
(\ref{56}) with multipolar expansions given by
(\ref{46},\ref{47},\ref{48},\ref{49},\ref{54},\ref{55}).

\section{Explicit Expansions for atoms with 1,2,3 electrons}

\subsection{Jacobi vectors}

As the center of mass vector $R$ and the $\left\{ r_{a}\right\}
_{1\leq a\leq N}$ are linearly dependent, one has to consider
Jacobi vectors $\left\{  R_{a}\right\} _{1\leq a\leq N}$ in order
to use independent variables. They are many such families of
vectors, each one being related to a definite building of the
center of mass which has to preserve the quantity
$I=\sum_{\alpha}m_{\alpha}r_{\alpha}^{2}=\sum_{\alpha}M_{\alpha
}R_{\alpha}^{2}$, where $\left(  \sum_{\alpha}m_{\alpha}\right)  R_{1}%
=\sum_{\alpha}m_{\alpha}r_{\alpha}$, $M_{1}=\left(
\sum_{\alpha}m_{\alpha }\right)  $ and $M_{\alpha}$ is the
effective mass corresponding to the associated $R_{\alpha}$.
Usually one proceeds imaging what is done for two particles,
pairing center of mass of (effective) particles getting at each
step the center of mass, the relative vector and the associated
masses.

\bigskip

\subsection{Octupole approximation for a neutral two body system}

We consider here the hydrogen atom. Here, $e_1=e,\, e_2=-e.$ We
introduce the Jacobi vectors $\{R_\alpha\}_{\alpha=1,2}$.

$\bigskip$
\\
We set
\begin{align*}
&R_2=r_{1}-r_{2},
\quad\frac{1}{M_2}=\frac{1}{m_{1}}+\frac{1}{m_{2}},\quad\quad
R_1=\frac{m_{1}r_{1}+m_{2}r_{2}}{M},\\ &M_1=m_{1}+m_{2},\,
P_1=-i\hbar\nabla_{R_1}, \quad P_2=-i\hbar\nabla_{R_2}.
\end{align*}%
We obtain
\[%
\begin{cases}
r_{1}=R_1+\frac{m_{2}}{M_1}R_2,\\ r_{2}=R_1-\frac{m_{1}}{M_1}R_2
\end{cases}
,\quad%
\begin{cases}
p_{1}=P_2+\frac{m_{1}}{M_1}P_1,\\ p_{2}=-P_2+\frac{m_{2}}{M_1}P_1.
\end{cases}
\]

In this new system of coordinates, $H_{mp}$ is given by:
\begin{equation}
H_{mp}=H_0+W_e,\label{57}
\end{equation}
where
\begin{eqnarray*}
H_0&=&\frac{P_1^2}{2M_1}+\frac{P_2^2}{2M_2}-\frac{e^2}{4\pi\varepsilon_0|R_2|}+H_f,\\
W_e&=&H_{mp}-H_0.
\end{eqnarray*}
Thus, letting $\mu=\alpha$ in $\left(\ref{46}-\ref{50}\right)$ the
octupole approximation of $\displaystyle\frac{1}{\mu^2}\Gamma
H_{mp}\Gamma^{-1}$ is given by: $\displaystyle H_0+W'_\mu,$\\
where
\begin{equation}
H_0=\frac{P_1^2}{2M_1}+\frac{P_2^2}{2M_2}-\frac{\hbar
c}{|R_2|}+H_f,\label{59}
\end{equation}
and
\begin{equation}
W'_\mu=\mu W^1_\mu+\mu^2 W^2_\mu+\mu^3 W^3_\mu+\mu^4
W^4_\mu,\label{60}
\end{equation}
with
\begin{equation}
\begin{split}
W^1_\mu&=-T_E^0(\mu)=-eR_2.E(\alpha R_1) ,\\
W^2_\mu&=-T_S^0(\mu)+T_M^0(\mu)+T_E^1(\mu)\\ &=-(M_1+M_2).B(\alpha
R_1)+\frac{1}{2M_1}[P_1.eR_2\wedge B(\alpha R_1)+h.c]\\
&+\frac{1}{4M_1}(1-\frac{4M_2}{M_1})^{\frac{1}{2}}[P_2.eR_2\wedge
B(\alpha R_1)+h.c]\\
&-\frac{1}{2}(1-\frac{4M_2}{M_1})^{\frac{1}{2}}eR_2.(R_2.\nabla_r)E(r)_{\mid_{r=\alpha
R_1}}, \\W^3_\mu&=T_M^1(\mu)+T_E^2(\mu)\\
&=\frac{1}{4M_1}(1-\frac{4M_2}{M_1})^{\frac{1}{2}}[P_1.eR_2\wedge(R_2.\nabla_r)B(r)_{\mid_{r=\alpha
R_1}}+h.c]\\
&+\frac{1}{6M_2}(1-\frac{3M_2}{M_1})[P_2.eR_2\wedge(R_2.\nabla_r)B(r)_{\mid_{r=\alpha
R_1}}+h.c]\\
&-\frac{1}{6}(1-\frac{3M_2}{M_1})eR_2.(R_2.\nabla_r)^2E(r)_{\mid_{r=\alpha
R_1}},\label{61.}
\end{split}
\end{equation}
\begin{equation}
\begin{split}
W^4_\mu&=T_M^2(\mu)+T_E^3(\mu)+T_{MM}^0(\mu)\\
&=\frac{1}{12M_1}(1-\frac{3M_2}{M_1})[P_1.eR_2\wedge(R_2.\nabla_r)^2B(r)_{\mid_{r=\alpha
R_1}}+h.c]\\&+\frac{M^3}{2M_2}(1-\frac{4M_2}{M_1})^{\frac{1}{2}}(1-\frac{2M_2}{M_1}
[P_2.eR_2\wedge(R_2.\nabla_r)^2B(r)_{\mid_{r=\alpha
R_1}}+h.c]\\&-\frac{1}{24}(1-\frac{4M_2}{M_1})^{\frac{1}{2}}(1-\frac{2M_2}{M_1})eR_2.(R_2.\nabla_r)^3
E(r)_{\mid_{r=\alpha R_1}}\\ &+\frac{1}{8M_2}(eR_2\wedge B(\alpha
R_1))^2.\label{61}
\end{split}
\end{equation}

Let us remark that, at the second order in $\mu$, the term
$\displaystyle R_2.(R_2.\nabla_r)E(r)_{\mid_{r=\mu R_1}}$ does not
appear in [GR] because the authors only consider the dipolar
approximation. The quadrupolar electric term appears as soon as
one considers the quadrupolar approximation.
\subsection{Quadrupole-approximation for a neutral three-body system}

Now we consider a neutral $3$ body system as, for example, the
helium atom $\displaystyle{He}^4$. Let $H_{mp}^{He}$ be the
Hamiltonian of the considered system. $\bigskip$

Consider a system formed of $3$ particles $A_1,A_2,A_3$ of
respective masses $m_1,m_2,$ $m_3,$ with charges
$e_1,\,e_2,\,e_3,$ of positions $r_1,\,r_2,\,r_3$ and of impulsion
$p_1,\,p_2,\,p_3.$

We introduce the Jacobi vectors associated to the following
partition: $a=\{(1,2),\,3\}.$ It is the case of a neutral atom
formed of a nucleus of charge $-2e$ and of two electrons each one
having a charge $e.$

We set

\begin{align*}
R_2  &  =r_{2}-r_{1},\
R_1=\frac{m_{1}r_{1}+m_{2}r_{2}+m_{3}r_{3}}{M_1},\
M_1=m_{1}+m_{2}+m_{3}\\
R_3&=r_{3}-\frac{m_{1}r_{1}}{M_2}-\frac{m_{2}r_{2}}{m},\
M_2=m_{1}+m_{2}.\\ P_1  &  =-i\hbar\nabla_{R_1},\
P_{2}=-i\hbar\nabla_{R_2},\ P_{3}=-i\hbar \nabla_{R_{3}}.
\end{align*}%
We obtain
\[%
\begin{cases}
r_{1}=R_1-\frac{m_{3}}{M_1}R_3-\frac{m_{2}}{M_2}R_{2}\\
r_{2}=R_1-\frac{m_{3}}{M_1}R_{3}+\frac{m_{1}}{M_2}R_{2}\\
r_{3}=R_1+\frac{M_2}{M_1}R_3%
\end{cases}
,\
\begin{cases}
p_{1}=\frac{m_{1}}{M_1}P_1-\frac{m_{1}}{M_2}P_3-P_2\\
p_{2}=\frac{m_{2}}{M_1}P_1-\frac{m_{2}}{M_2}P_3+P_2\\
p_{3}=\frac{m_{3}}{M_1}P_1+P_3+P_2%
\end{cases}
\]

In the case of the Helium atom we have $m_1=m_2,\,e_1=e_2=e$ and
$e_3=-2e.$\\ In this new system of coordinates, $H_{mp}^{He}$ is
given by:

\begin{equation*}
\begin{split}
H_{mp}^{He}&=\frac{1}{2m_1}
\big[\frac{m_1}{M_1}P_1-\frac{1}{2}P_3-P_2
+\int_{\er^3}\Theta_1(r)\wedge B(r)d^3r\big]^2\\ &+\frac{1}{2m_1}
\big[\frac{m_1}{M_1}P_1-\frac{1}{2}P_3+P_2
+\int_{\er^3}\Theta_2(r)\wedge B(r)d^3r\big]^2\\ &+\frac{1}{2m_3}
\big[\frac{m_3}{M_1}P_1+P_3+\int_{\er^3}\Theta_3(r)\wedge
B(r)d^3r\big]^2\\ &+
\frac{e^2}{4\pi\varepsilon_0|R_2|}-\frac{2e^2}{4\pi\varepsilon_0|R_3+\frac{1}{2}R_2|}
-\frac{2e^2}{4\pi\varepsilon_0|R_3-\frac{1}{2}R_2|}\\
&-\sum_{\alpha=1}^3M_\alpha.B(r_\alpha)+H_f-\int_{\er^3}E(r).P(r)d^3r.
\end{split}
\end{equation*}
We write $H_{mp}^{He}$ under the following form:
$$H_{mp}^{He}=H_0^{He}+W_e,$$ with
\begin{equation}
\begin{split}
H_0^{He}:&=\frac{P_1^2}{2M_1}+\frac{P_3^2}{2\mu_{12,3}}+\frac{P_2^2}{2m_1}
+\frac{e^2}{4\pi\varepsilon_0|R_2|}-\frac{2e^2}{4\pi\varepsilon_0|R_3+\frac{1}{2}R_2|}\\
&-\frac{2e^2}{4\pi\varepsilon_0|r_3-\frac{1}{2}R_2|}+H_f,
\label{62}
\end{split}
\end{equation}
and $$W_e:=H_{mp}^{He}-H_0^{He}.$$ Here
$\frac{1}{\mu_{12,3}}=\frac{1}{2m_1}+\frac{1}{m_3}.$

Letting $N=3$ in $(\ref{46}-\ref{50})$ we get at the second order
in $\mu=2\alpha$:
\begin{equation}
H_{\mu}^Q=\frac{1}{\mu^2}\Gamma
H_{mp}^{He}\Gamma^{-1}=H_0^{He}+W_\mu,\label{63}
\end{equation}
where,
\begin{equation}
H_0^{He}=\frac{P_1^2}{2M_1}+\frac{P_3^2}{2\mu_{12,3}}+\frac{P_2^2}{m_1}+\frac{\hbar
c}{2|R_2|}-\frac{\hbar c}{|R_3+\frac{1}{2}R_2|}-\frac{\hbar
c}{|R_3-\frac{1}{2}R_2|}+H_f,\label{64}
\end{equation}
\begin{equation}
W_\mu=\mu W^1_\mu+\mu^2W^2_\mu,\label{65}
\end{equation}
with
\begin{equation}
\begin{split}
W^1_\mu&=-T_E^0(\mu)=2eR_3.E(2\alpha R_1),\\
W^2_\mu&=-T_S^0(\mu)+T_M^0(\mu)+T_E^1(\mu)\\
&=-(M_1+M_2+M_3).B(2\alpha R_1)-\frac{1}{M_1}[P_1.eR_3\wedge
B(2\alpha R_1)+h.c]\\
&+\frac{1}{2\mu_{12,3}}(1-\frac{4\mu_{12,3}}{M_1})^{\frac{1}{2}}[P_3.eR_3\wedge
B(2\alpha R_1)+h.c]\\ &+\frac{1}{4m_1}[P_2.eR_2\wedge B(2\alpha
R_1)+h.c]\\
&-e[(1-\frac{4\mu_{12,3}}{M_1})^{\frac{1}{2}}R_3.(R_3.\nabla_r)+\frac{1}{4}
R_2.(R_2.\nabla_r)]E(r)_{\mid_{r=2\alpha R_1}}. \label{66}
\end{split}
\end{equation}
\subsection{Quadrupole-approximation for a neutral four-body system}
We are finally interested in the case of neutral moving system
formed of four bodies. For this system we consider the following
partition for $\{1,2,3,4\}:$ $$c=\{a,4\},\,a=\{1,2,3\}.$$

 This partition correspond to the lithium atom.
\\
We define the Jacobi vectors as follows:
\begin{eqnarray*}
R_2&=&r_1-r_2,\quad \quad R_3=r_3-\frac{m_1r_1+m_2r_2}{m_1+m_2},
\quad R_4=(R_4,R_3)\\
\\
R_4&=&r_4-\frac{m_1r_1+m_2r_2+m_3r_3}{m_1+m_2+m_3},\quad
M_4=m_1+m_2+m_3\\
\\
R_1&=&\frac{m_1r_1+m_2r_2+m_3r_3+m_4r_4}{M},\, M_1=m_1+m_2+m_3+m_4
\end{eqnarray*}
\begin{eqnarray*}
P_2&=&-i\hbar\nabla_{R_2},\quad P_3=-i\hbar\nabla_{R_3},\quad
P_4=-i\hbar\nabla_{R_4}, \quad P_1=-i\hbar\nabla_{R_1}.
\end{eqnarray*}
We suppose here that $m_1=m_2=m_3$ and we set $m_1=m_2=m_3$ and
$m_4=m_2.$

We then obtain
$$
\begin{cases}
r_1=R_1-\frac{m_2}{M_1}R_4-\frac{1}{3}R_3+\frac{1}{2}R_2\\
r_2=R_1-\frac{m_2}{M_1}R_4-\frac{1}{3}R_3-\frac{1}{2}R_2\\
r_3=R_1-\frac{m_2}{M_1}R_4-\frac{2}{3}R_3\\
r_4=R_1+\frac{m_a}{M_1}R_4
\end{cases},\
\begin{cases}
p_1=\frac{m_1}{M_1}P_1-\frac{1}{3}P_4-\frac{1}{2}P_3+P_2\\
p_2=\frac{m_1}{M_1}P_1-\frac{1}{3}P_4-\frac{1}{2}P_3-P_2\\
p_3=\frac{m_1}{M_1}P_1-\frac{1}{3}P_4+P_3\\
p_4=\frac{m_2}{M_1}P_1+P_4.
\end{cases}
$$ We recalls that the Hamiltonian $H_{mp}^{Li}$ of the considered
system is given by:
\begin{eqnarray*}
H_{mp}^{Li}&=&\sum_{\alpha=1}^4\frac{1}{2m_\alpha}
\big(p_\alpha+\int d^3r\Theta_\alpha(r)\wedge B(r)\big)^2
+\sum_{1\leq\alpha<\beta\leq4}\frac{e_\alpha e_\beta}
{4\pi\varepsilon_0|r_\alpha-r_\beta|}\\
&-&\sum_{\alpha=1}^4M_\alpha.B(r_\alpha) +\frac{1}{2}\int
d^3r\big(\frac{(\Pi(r)+P(r))^2}{\varepsilon_0}
+\frac{B(r)^2}{\mu_0}\big).
\end{eqnarray*}
We now consider that $e_1=e_2=e_3=e$ and $e_4=-3e,$ the lithium
atom.

In the new system of coordinates the Hamiltonian $H_{mp}^{Li}$ is
written as follows:
\begin{eqnarray*}
H_{mp}^{Li}&=&\frac{1}{2m_1}
\big(\frac{m_1}{M_1}P_1-\frac{1}{3}P_4-\frac{1}{2}P_3+P_2 +\int
d^3r\Theta_1(r)\wedge B(r)\big)^2\\
\\
&+&\frac{1}{2m_1}\big(\frac{m_1}{M_1}P_1-\frac{1}{3}P_4-\frac{1}{2}P_3-P_2
+\int d^3r\Theta_2(r)\wedge B(r)\big)^2\\
\\
&+&\frac{1}{2m_1}\big(\frac{m_1}{M_1}P_1-\frac{1}{3}P_4+P_3 +\int
d^3r\Theta_3(r)\wedge B(r)\big)^2\\
\\
&+&\frac{1}{2m_2}\big(\frac{m_2}{M_1}P_1+P_4 +\int
d^3r\Theta_4(r)\wedge B(r)\big)^2\\
\\
&+&\frac{e^2}{4\pi\v_0|R_2|}+\frac{e^2}{4\pi\v_0|R_3-\frac{1}{2}R_2|}
+\frac{e^2}{4\pi\v_0|R_3+\frac{1}{2}R_2|}\\
\\
&-&\frac{3e^2}{4\pi\v_0|R_4+\frac{1}{3}R_3-\frac{1}{2}R_2|}
-\frac{3e^2}{4\pi\v_0|R_4+\frac{1}{3}R_3+\frac{1}{2}R_2|}\\
\\
&-&\frac{3e^2}{4\pi\v_0|R_4-\frac{2}{3}R_3|}
+H_f-\sum_{\alpha=1}^4M_\alpha.B(r_\alpha)-\int E(r).P(r)d^3r.
\end{eqnarray*}
We then write under the following form:
$$H_{mp}^{Li}=H_0^{Li}+W_e,$$ where
\begin{equation}
\begin{split}
H_{0}^{Li}&=\frac{P^2_1}{2M_1}+\frac{M_1}{6m_1m_2}({P_4})^2
+\frac{3}{4m_1}(P_3)^2+\frac{1}{m_1}{P_2}^2\\
&+\frac{e^2}{4\pi\v_0|R_2|}+\frac{e^2}{4\pi\v_0|R_3-\frac{1}{2}R_2|}
+\frac{e^2}{4\pi\v_0|R_3+\frac{1}{2}R_2|}\\
&-\frac{3e^2}{4\pi\v_0|R_4+\frac{1}{3}R_3-\frac{1}{2}R_2|}
-\frac{3e^2}{4\pi\v_0|R_4+\frac{1}{3}R_3+\frac{1}{2}R_2|}\\
&-\frac{3e^2}{4\pi\v_0|R_4-\frac{2}{3}R_3|}+H_f,
\end{split}
\end{equation}
and
\begin{equation}
\begin{split}
W_e&=-\sum_{\alpha=1}^4M_\alpha.B(R_1)-\int E(r).P(r)d^3r\\
&+\frac{1}{2M_1}\big(P_1.\int d^3r(\sum_{j=1}^4\Theta_j(r)\wedge
B(r)+ \int d^3r(\sum_{j=1}^4\Theta_j(r)\wedge B(r).P_1\big)\\
&-\frac{1}{6m_1}\big(P_4.\int d^3r(\sum_{j=1}^3\Theta_j(r)\wedge
B(r)+ \int d^3r(\sum_{j=1}^3\Theta_j(r)\wedge B(r).P_4\big)\\
&+\frac{1}{2m_2}\big(P_4.\int d^3r\Theta_4(r)\wedge B(r) +\int
d^3r\Theta_4(r)\wedge B(r).P_4\big)\\
&-\frac{1}{4m_1}\big(P_3.\int d^3r(\Theta_1(r)+\Theta_2(r)
-2\Theta_3(r))\wedge B(r)\\ &+\int d^3r(\Theta_1(r)+\Theta_2(r)
-2\Theta_3(r))\wedge B(r).P_3\big)\\ &+\frac{1}{2m_1}\big(P_2.\int
d^3r(\Theta_1(r)-\Theta_2(r))\wedge B(r)
\\
&+\int d^3r(\Theta_1(r)-\Theta_2(r))\wedge B(r).P_2\big)\\
&+\frac{1}{2m_1}\big\{\big(\int d^3r\Theta_1(r)\wedge B(r)\big)^2
+\big(\int d^3r\Theta_2(r)\wedge B(r)\big)^2\big\}\\
&+\frac{1}{2m_1}\big(\int d^3r\Theta_3(r)\wedge B(r)\big)^2
+\frac{1}{2m_2}\big(\int d^3r\Theta_4(r)\wedge B(r)\big)^2.
\end{split}
\end{equation}
Let
\begin{equation}
H^{Li}_{\mu}=\frac{1}{\mu^2}\Gamma H_{mp}^{Li}\Gamma^{-1}.
\end{equation}
Setting $N=4$ in $(\ref{46}-\ref{50})$ we obtain at the second
order in $\mu=3\alpha$:
\begin{equation}
H^{Li}_{\mu}=H_0^{Li}+\mu W^1_\mu+\mu^2 W^2_\mu,
\end{equation}
with
\begin{equation}
\begin{split}
H_0^{Li}&=\frac{P_1^2}{2M_1}+\frac{M_1}{6m_1m_2}{P_4}^2
+\frac{3}{4m_1}{P_3}^2+\frac{1}{m_1}{P_2}^2\\ &+\frac{\hbar
c}{|R_2|}+\frac{\hbar c}{|R_3-\frac{1}{2}R_2|} +\frac{\hbar
c}{|R_3+\frac{1}{2}R_2|}\\ &-\frac{3\hbar
c}{|R_4+\frac{1}{3}R_3-\frac{1}{2}R_2|} -\frac{3\hbar
c}{|R_4+\frac{1}{3}R_3+\frac{1}{2}R_2|} -\frac{3\hbar
c}{|R_4-\frac{2}{3}R_3|}+H_f,
\end{split}
\end{equation}

\begin{equation}
\begin{split}
W^1_\mu&=-T_E^0(\mu)=-3eR_4.E(3\alpha R_1),\\
W^2_\mu&=-T_S^0(\mu)+T_M^0(\mu)+T_E^1(\mu)\\
&=-[M_1+M_2+M_3+M_4].B(3\alpha R_1)\\
&+\big[\frac{3(m_2-3m_1)e}{2M_1}R_4.(R_4.\nabla_r)
+\frac{e}{3}R_3.(R_3.\nabla_r)+\frac{e}{4}R_2.(R_2.\nabla_r)\big].\\
&E(r) _{\mid_{r=3\alpha R_1}}\\ &+\frac{1}{2M_1}[P_1.d\wedge
B(3\alpha R_1)+h.c] +\frac{1}{12}[P_4.d\wedge B(3\alpha
R_1)+h.c]\\ &+\frac{1}{4m_1}[P_3.eR_3\wedge B(3\alpha R_1)+h.c]
+\frac{1}{4m_1}[P_2.eR_2\wedge B(3\alpha R_1)+h.c].
\end{split}
\end{equation}

We observe that the several approximations for the neutral
composite systems of $2,3$ and $4$ particles we have obtained are
exactly the same as the ones got by [F]. In [F] the author first
computes the multipolar expansion and then implements the scaling
analysis.
\section{Conclusion:}
We have found again (see [GR]) the electric dipolar term at first
order in $\mu$ and at second order in $\mu$ the magnetic dipolar
term together with the R\"ontgen one that already appear in the
dipolar approximation of $H_{mp}.$ But now, considering the full
multipolar expansion in $\mu,$ we have shown that an extra term,
i.e., an electric quadrupolar term appears also at the second
order in $\mu.$

This strongly suggests that we must finally consider in energy
computation the following Hamiltonian instead of the usual dipolar
one:
\begin{equation}
H_{mp}^{(2)}=H_0+(-T_E^0+\{-T_S^0+T_M^0+T_E^1\}).
\end{equation}
As the symmetries of the involved terms are different, a new level
of complexity has to be taken into account. For instance for the
computation of the spontaneous emission of these atomic system,
the deductions of [WB] have to be consider again.

\newpage
\pn {\bf\Large References}

\vspace{1cm}
\pn {[BBB]} {\bf L. G. Boussiakou, C. R. Bennett and M. Babiker.}
{\sl} Phys. Rev. Lett., $\underline{89}$, $123001$ (2002).

\vspace{0.3cm}

\pn {[BFS]} {\bf V. Bach, J. Fr\"ohlich and M. Sigal}. {\sl }
Adv.Math., $\underline{137}$, p.299-395 (1998).

\vspace{0.3cm}

\pn {[F]} {\bf A. Felhi}. {\sl } Ph. D.  universit{\'e}
Paris-Nord. (2002).

\vspace{0.3cm}

\pn {[FFG]} {\bf C. Fefferman, J. Fr\"ohlich and G.M. Graf}. {\sl}
 Com. math. Phys., $\underline{190}$, p.309-330, (1997).

\vspace{0.3cm}

\pn {[GR]} {\bf J.-C. Guillot and J. Robert.} {\sl} J. Phys. A:
Math. Gen. $\underline{35},$ p.5023-5039.

\vspace{0.3cm}

\pn {[H]} {\bf W. P. Healy.} J. Phys. A: Math. Gen.
$\underline{10}$ No $2$ (February $1977$) p.279-298

\vspace{0.3cm}

\pn {[LBBL]} {\bf V. Lembessis, M. Babiker, C. Baxter and R.
London.} Phys. Rev. A., $\underline{48}$, p.1594 (1993).

\vspace{0.3cm}

\pn {[M]} {\bf A. Messiah.} {\sl M{\'e}canique quantique.} Dunod,
Paris (1995).

\vspace{0.3cm}

\pn {[S]} {\bf S.S. Schweber.} {\sl An Introduction to
Relativistic Quantum Field Theory.} Row, Peterson and Company,
(1961).

\vspace{0.3cm}

\pn {[WB]} {\bf M. Wilkens and M. Babiker.} Phys. Rev. A.,
$\underline{48}$, p.570, (1994) .

\end{document}